\documentclass[usegraphicx]{mn2e}

\title[A Toomre-like stability criterion for the clumpy and turbulent
       interstellar medium]
      {A Toomre-like stability criterion for the clumpy and turbulent
       interstellar medium}

\author[A. B. Romeo, A. Burkert and O. Agertz]
       {Alessandro B. Romeo,$^{1}$\thanks{E-mail: romeo@chalmers.se}
        Andreas Burkert$^{2}$ and
        Oscar Agertz$^{3}$\\
        $^{1}$Onsala Space Observatory,
              Chalmers University of Technology,
              SE-43992 Onsala, Sweden\\
        $^{2}$University Observatory,
              University of Munich,
              Scheinerstr. 1, D-81679 Munich, Germany\\
        $^{3}$Institute for Theoretical Physics,
              University of Z\"{u}rich,
              CH-8057 Z\"{u}rich, Switzerland}

\begin{document}

\date{Accepted 2010 May 6.
      Received 2010 April 22; in original form 2010 January 26}

\pagerange{\pageref{firstpage}--\pageref{lastpage}}

\pubyear{2010}

\maketitle

\label{firstpage}

\begin{abstract}
We explore the gravitational instability of clumpy and turbulent gas discs,
taking into account the Larson-type scaling laws observed in giant molecular
clouds (GMCs) and H\,\textsc{i}, as well as more general scaling relations.
This degree of freedom is of special interest in view of the coming high-$z$
ISM surveys, and is thus potentially important for understanding the
dynamical effects of turbulence at all epochs of galaxy evolution.  Our
analysis shows that turbulence has a deep impact on the gravitational
instability of the disc.  It excites a rich variety of stability regimes,
several of which have no classical counterpart.  Among other diagnostics, we
provide \emph{two useful tools} for observers and simulators: (1) the
stability map of turbulence, which illustrates our stability scenario and
relates it to the phenomenology of interstellar turbulence: GMC/H\,\textsc{i}
observations, simulations and models; (2) a Toomre-like stability criterion,
$Q\geq\overline{Q}$, which applies to a large class of clumpy/turbulent
discs.  We make specific predictions about GMC and cold-H\,\textsc{i}
turbulence, and point out the implications of our analysis for high-$z$
galaxy surveys.
\end{abstract}


\section{INTRODUCTION}

Toomre's (1964) stability criterion, $Q\geq1$, is one of the pillars of disc
galaxy dynamics (see, e.g., Binney \& Tremaine 2008).  It is used in a wide
variety of contexts; in star formation for example, where the gravitational
instability of the interstellar gas plays a critical role (e.g., Quirk 1972;
Kennicutt 1989; Martin \& Kennicutt 2001; Schaye 2004, 2008; Burkert 2009;
Elmegreen 2009).

   So why introduce a new stability criterion?  Because behind Toomre's
criterion there is one fundamental assumption: the medium is approximately in
equilibrium, with well-defined surface density $\Sigma$ and velocity
dispersion $\sigma$.  But this is far from being true in the clumpy and
turbulent interstellar gas, where such quantities depend strongly on $\ell$,
the size of the region over which they are measured.  In fact, a fundamental
aspect of interstellar turbulence is the existence of density- and
velocity-size scaling laws: $\Sigma\propto\ell^{a}$ and
$\sigma\propto\ell^{b}$ (see, e.g., Elmegreen \& Scalo 2004; McKee \&
Ostriker 2007).  In giant molecular clouds (GMCs), the scaling exponents are
$a\approx0$ and $b\approx\frac{1}{2}$ (e.g., Larson 1981; Solomon et al.\
1987; Bolatto et al.\ 2008; Heyer et al.\ 2009; Hughes et al.\ 2010).  In the
H\,\textsc{i} component, density and velocity fluctuations seem to have a
Kolmogorov spectrum \emph{up to galactic scales}: $a\sim\frac{1}{3}$ for
$\ell\la\mbox{10 kpc}$, and $b\sim\frac{1}{3}$ for $\ell\la\mbox{1 kpc}$
(e.g., Lazarian \& Pogosyan 2000; Elmegreen et al.\ 2001; Begum et al.\ 2006;
Kim et al.\ 2007; Dutta et al.\ 2008; Roy et al.\ 2008; Dutta et al.\
2009a,\,b).  Note, however, that the uncertainties are large, especially in
the H\,\textsc{i} case.  Further evidence for ISM turbulence is provided by
the `Big Power Law in the Sky', i.e.\ the fact that electron density
fluctuations show a Kolmogorov spectrum over a wide range of scales: from
$10^{3}$ km up to 10 pc (Armstrong et al.\ 1995; see Chepurnov \& Lazarian
2010 for the most recent determination of the upper scale).

   Clumpy and turbulent gas is also observed in high-redshift star-forming
galaxies, where it dominates the morphology and dynamics of the disc (e.g.,
Wadadekar et al.\ 2006; Elmegreen et al.\ 2007; Genzel et al.\ 2008; Shapiro
et al.\ 2008; F\"{o}rster Schreiber et al.\ 2009).  Coming surveys will tell
us how $\Sigma$ and $\sigma$ scale with $\ell$ at high $z$, and thus how
turbulence develops in disc galaxies.  This is one of the hot topics in
modern astrophysics (e.g., Krumholz \& Burkert 2010), and several
state-of-the-art simulations have already been designed for such a purpose
(e.g., Wada et al.\ 2002; Kim \& Ostriker 2007; Levine et al.\ 2008; Agertz
et al.\ 2009a,\,b; Bournaud \& Elmegreen 2009; Dekel et al.\ 2009; Tasker \&
Tan 2009).  In order to interpret such data correctly, one must understand in
detail how turbulence affects the (in)stability of the disc.

   \emph{These facts all together motivate a thorough investigation of the
problem.}

   Elmegreen (1996) assumed Larson-type scaling relations,
$\Sigma\propto\ell^{a}$ and $\sigma\propto\ell^{b}$, and investigated the
case $a=-1$ and $b=\frac{1}{2}$.  He found that the disc is always stable at
large scales and unstable at small scales.  To the best of our knowledge,
this was the only theoretical work devoted to the gravitational instability
of clumpy/turbulent discs before ours.  In contrast, several investigations
focused on the effects of turbulence on Jeans instability (e.g., Bonazzola et
al.\ 1987; Just et al.\ 1994; V\'{a}zquez-Semadeni \& Gazol 1995).

   The goal of our paper is to explore the gravitational instability of
clumpy/turbulent discs, spanning the whole range of values for $a$ and $b$.
Among other diagnostics, we provide two useful tools for observers and
simulators:
\begin{enumerate}
\item \emph{The stability map of turbulence.}  Using $a$ and $b$ as
      coordinates, we illustrate the natural variety of stability regimes
      possessed by such discs, and populate this diagram with observations,
      simulations and models of interstellar turbulence (Fig.\ 1).
\item \emph{A Toomre-like stability criterion.}  We show that in our map
      there is a densely populated domain where the stability criterion is of
      the form $Q\geq\overline{Q}$, and determine the stability threshold
      $\overline{Q}$ as a function of $a$, $b$ and of the scale at which $Q$
      is measured [Eq.\ (17) and Fig.\ 2].  If our criterion is fulfilled,
      then the disc is stable at all scales (the case investigated by
      Elmegreen 1996 lies in another stability regime).
\end{enumerate}

   The rest of the paper is organized as follows.  In Sect.\ 2, we determine
a dispersion relation that takes fully into account the scaling laws of
interstellar turbulence as well as the thickness of the gas layer.  We then
surf through the various stability regimes.  Our Toomre-like stability
criterion is given in Sect.\ 2.7, together with other important diagnostics:
the most unstable scale and its growth rate.  In Sect.\ 3, we discuss the
stability map of turbulence.  We then make specific predictions about
strongly and weakly supersonic turbulence.  In Sect.\ 4, we explore the
densely populated Toomre-like domain and illustrate the stability diagnostics
in a number of cases, taking into account the saturation of density and
velocity at large scales.  In Sect.\ 5, we draw the conclusions.

\section{STABLE OR UNSTABLE ?}

\subsection{The turbulent dispersion relation}

Consider a gas disc of scale height $h$, and perturb it with axisymmetric
waves of frequency $\omega$ and wavenumber $k$.  The response of the disc is
described by the dispersion relation
\begin{equation}
\omega^{2}=\kappa^{2}-\frac{2\pi G\Sigma\,k}{1+kh}+\sigma^{2}\,k^{2}\,,
\end{equation}
where $\kappa$ is the epicyclic frequency, $\Sigma$ is the surface density at
equilibrium, and $\sigma$ is the sound speed (Romeo 1990, 1992, 1994; see
also Vandervoort 1970).  So the three terms on the right side of Eq.\ (1)
represent the contributions of rotation, self-gravity and pressure.  For
$kh\ll1$, Eq.\ (1) reduces to the usual dispersion relation for an
infinitesimally thin gas disc (see, e.g., Binney \& Tremaine 2008).  For
$kh\gg1$, one recovers the case of Jeans instability with rotation, since
$\Sigma/h\approx2\rho$.  In other words, scales comparable to $h$ mark the
transition from 2D to 3D stability.  Note that we can encapsulate the effect
of thickness in a single quantity and rewrite Eq.\ (1) as
\begin{equation}
\omega^{2}=\kappa^{2}-2\pi G\Sigma_{\mathrm{eff}}\,k+\sigma^{2}\,k^{2}\,,
\end{equation}
where $\Sigma_{\mathrm{eff}}=\Sigma/(1+kh)$ is the effective surface density.
$\Sigma_{\mathrm{eff}}$ and $\sigma$ are two important quantities, which we
discuss in detail below.

\begin{description}
\item \emph{The effective surface density.}  What is the relation between
      $\Sigma_{\mathrm{eff}}$ and $\rho$?  Since
      $\Sigma_{\mathrm{eff}}=\Sigma/(1+kh)$ and $\Sigma\sim\rho h$, we find
      that $\Sigma_{\mathrm{eff}}\sim\rho h$ if $kh\la1$, and
      $\Sigma_{\mathrm{eff}}\sim\rho/k$ otherwise.  This means that the
      observational counterpart of $\Sigma_{\mathrm{eff}}$ is the mass column
      density measured over a region of size $\ell=1/k$.  In the cold ISM,
      which is highly supersonic and hence strongly compressible, the
      amplitude of density fluctuations is typically much larger than the
      mean density.  Density fluctuations have a power-law energy spectrum,
      $E_{\rho}(k)\propto k^{-r}$, which means $\rho\propto k^{-(r-1)/2}$
      (see, e.g., Elmegreen \& Scalo 2004).  A power-law spectrum is then
      imprinted on $\Sigma_{\mathrm{eff}}$:
      \begin{equation}
      \Sigma_{\mathrm{eff}}=\Sigma_{0}\left(\frac{k}{k_{0}}\right)^{-a}\,.
      \end{equation}
      Using the relation between $\Sigma_{\mathrm{eff}}$ and $\rho$ that we
      have found above, we can relate $a$ to $r$: $a=\frac{1}{2}\,(r-1)$ for
      $kh\la1$, while $a=\frac{1}{2}\,(r+1)$ for $kh\ga1$.  In the warm ISM,
      which is transonic or subsonic and hence weakly compressible, the
      density contrast is typically small so $\Sigma_{\mathrm{eff}}$ is
      dominated by the mean density, as in the limiting case of a
      non-turbulent disc: $\Sigma_{\mathrm{eff}}=constant$ for small $k$,
      while $\Sigma_{\mathrm{eff}}\propto k^{-1}$ for large $k$.  Hereafter
      we will omit the subscript `eff', unless otherwise stated.
\item \emph{The sound speed.}  The observational counterpart of $\sigma$ is
      the 1D velocity dispersion measured over a region of size $\ell=1/k$:
      $\sigma^{2}= \sigma_{\mathrm{ther}}^2+\sigma_{\mathrm{tur}}^{2}(k)$,
      where $\sigma_{\mathrm{ther}}$ and $\sigma_{\mathrm{tur}}$ are the
      thermal and turbulent 1D velocity dispersions.  Velocity fluctuations
      have a power-law energy spectrum, $E_{v}(k)\propto k^{-s}$, which means
      $\sigma_{\mathrm{tur}}\propto k^{-(s-1)/2}$ (see, e.g., Elmegreen \&
      Scalo 2004).  Both thermal and turbulent motions tend to support the
      gas against gravitational instability.  However, as pointed out by the
      referee, turbulent support results in relations that are true only in a
      statistical sense, with individual clouds collapsing even when the
      statistical criterion for stability is well satisfied.  In the cold
      ISM, $\sigma$ has a power-law dependence on $k$ since it is dominated
      by $\sigma_{\mathrm{tur}}(k)$:
      \begin{equation}
      \sigma=\sigma_{0}\left(\frac{k}{k_{0}}\right)^{-b}\,,
      \end{equation}
      where $b=\frac{1}{2}\,(s-1)$ and $s$ is the velocity spectral index.
      In the warm ISM, $\sigma(k)$ deviates significantly from a power law
      since $\sigma_{\mathrm{ther}}$ is no longer negligible.
\end{description}

   Hereafter we will only consider the \emph{cold} ISM ($\mathrm{H}_{2}$ and
cold H\,\textsc{i}), since in such a case we can explicitely take into
account the power-law scaling of interstellar turbulence via Eqs (3) and (4).
We will regard the quantity $\ell_{0}=1/k_{0}$ introduced in those equations
as the fiducial scale at which the mass column density and the 1D velocity
dispersion are observed.  This is also the scale at which $Q$ and other
stability quantities are measured.  Substituting Eqs (3) and (4) into Eq.\
(2), we obtain the final dispersion relation:
\begin{equation}
\omega^{2}=\kappa^{2}-2\pi G\Sigma_{0}k_{0}^{a}\,k^{1-a}+\sigma_{0}^{2}k_{0}^
{2b}\,k^{2(1-b)}\,.
\end{equation}

   Our phenomenological approach differs significantly from the traditional
way to include turbulent effects in the dispersion relation, which is to
identify $\sigma$ with the typical 1D velocity dispersion observed at
galactic scales.  Bonazzola et al.\ (1987), Just et al.\ (1994),
V\'{a}zquez-Semadeni \& Gazol (1995) and Elmegreen (1996) adopted an approach
similar to ours for investigating the gravitational instability of
clumpy/turbulent media.

   Eq.\ (5) is the starting-point of our stability analysis.  As in the usual
case, stability at all scales requires that $\omega^{2}\geq0$ for all $k$.%
\footnote{The dispersion relation assumes that $kR\gg1$, where $R$ is the
          radial coordinate (see, e.g., Binney \& Tremaine 2008).  This local
          condition is more restrictive than the natural requirement
          $k\ga2\pi/L$, where $L$ is the size of the gas disc.  Since the
          condition above cannot be rigorously included in the stability
          analysis, the usual procedure is to consider all $k$ and to check a
          posteriori whether $kR\gg1$ or not.}
Whether this requirement can be satisfied or not depends on the self-gravity
and pressure terms, which now scale as $k^{A}$ and $k^{B}$:
\begin{equation}
A=1-a\,,
\end{equation}
\begin{equation}
B=2\left(1-b\right)\,.
\end{equation}
In the following, we explore all the various cases.

\subsection{Case $a=1,\;b\neq1$}

Here the self-gravity term is independent of $k$ ($A=0$), like the rotation
term:
\begin{equation}
\omega^{2}=\underbrace{\kappa^{2}-2\pi G\Sigma_{0}k_{0}}_{\textstyle C}\mbox{
}+\sigma_{0}^{2}k_{0}^{2b}\,k^{2(1-b)}\,.
\end{equation}
For $b<1$, $\omega^{2}$ increases with $k$ and tends to $C$ as
$k\rightarrow0$.  Hence the sign of $C$ determines whether the disc is stable
at all scales or not.  Stability requires that $C\geq0$.  For $b>1$,
$\omega^{2}$ decreases with $k$ and tends to $C$ as $k\rightarrow\infty$.
So, again, stability requires that $C\geq0$.  The stability criterion is
then:
\begin{equation}
\mbox{STABLE\ }\forall k\Longleftrightarrow
k_{0}\leq k_{\mathrm{T}}=\frac{\kappa^{2}}{2\pi G\Sigma_{0}}\,.
\end{equation}
Eq.\ (9) resembles Toomre's stability criterion for cold discs, $k\leq
k_{\mathrm{T}}$, but there is one important difference: Eq.\ (9) is a
condition that must be fulfilled by $k_{0}$ to ensure stability for all $k$,
whereas the cold Toomre criterion ensures stability for small $k$.

\subsection{Case $a\neq1,\;b=\frac{1}{2}\,(1+a)$}

Here the pressure term has the same $k$-dependence as the self-gravity term
($B=A$):
\begin{equation}
\omega^{2}=\kappa^{2}+\underbrace{\left(\sigma_{0}^{2}k_{0}^{1+a}-2\pi G
\Sigma_{0}k_{0}^{a}\right)}_{\textstyle C}\,k^{1-a}\,.
\end{equation}
If $a<1$, then $\omega^{2}$ converges to $\kappa^{2}$ as $k\rightarrow0$ and
the sign of $\mathrm{d}\omega^{2}/\mathrm{d}k$ is equal to the sign of $C$
(for $C<0$, $\omega^{2}$ diverges as $k\rightarrow\infty$).  Hence stability
requires that $C\geq0$.  If $a>1$, then $\omega^{2}$ converges to
$\kappa^{2}$ as $k\rightarrow\infty$ and the sign of
$\mathrm{d}\omega^{2}/\mathrm{d}k$ is opposite to the sign of $C$ (for $C<0$,
$\omega^{2}$ diverges as $k\rightarrow0$).  So, again, stability requires
that $C\geq0$.  The stability criterion is then:
\begin{equation}
\mbox{STABLE\ }\forall k\Longleftrightarrow
k_{0}\geq k_{\mathrm{J}}=\frac{2\pi G\Sigma_{0}}{\sigma_{0}^{2}}\,.
\end{equation}
The resemblance between Eq.\ (11) and the 2D Jeans criterion, $k\geq
k_{\mathrm{J}}$, is superficial.  In fact, Eq.\ (11) ensures that the disc is
stable at all, rather than small, scales.  An analogous fact was noted in the
context of our first stability criterion [Eq.\ (9)].

\subsection{Case $a=1,\;b=1$}

Although this case seems the intersection of Case 2.2 and Case 2.3, the
stability criterion is not $k_{\mathrm{J}}\leq k_{0}\leq k_{\mathrm{T}}$.  In
fact, this case is singular.  The self-gravity and pressure terms are
independent of $k$ ($A=B=0$), like the rotation term, so $\omega^{2}(k)$ is
constant:
\begin{equation}
\omega^{2}=\underbrace{\kappa^{2}-2\pi G\Sigma_{0}k_{0}+\sigma_{0}^{2}k_{0}^{
2}}_{\textstyle C}\,.
\end{equation}
As $C$ is quadratic in $k_{0}$, the inequality $C\geq0$ can be easily solved
and the resulting stability criterion is:
\begin{equation}
\mbox{STABLE\ }\forall k\Longleftrightarrow
\left\{\begin{array}{ll}
       k_{0}\leq k_{-}\mbox{\ or\ }k_{0}\geq k_{+} & \mbox{if\ }Q<1\,, \\
       0<k_{0}<\infty                              & \mbox{else}\,.
       \end{array}
\right.
\end{equation}
Here $k_{-}$ and $k_{+}$ are related to the Toomre wavenumber
$k_{\mathrm{T}}$ (and to the Jeans wavenumber
$k_{\mathrm{J}}=k_{\mathrm{T}}\,4/Q^{2}$):
\begin{equation}
k_{\pm}=k_{\mathrm{T}}\,\frac{2}{Q^{2}}\left(1\pm\sqrt{1-Q^{2}}\right)\,;
\end{equation}
and $Q$ is the Toomre parameter:
\begin{equation}
Q=\frac{\kappa\sigma_{0}}{\pi G\Sigma_{0}}\,.
\end{equation}
The $Q<1$ case of Eq.\ (13) resembles the corresponding Toomre stability
condition, but see the remarks following Eqs (9) and (11).  In contrast, the
$Q\geq1$ case is identical to Toomre's criterion.

\subsection{Case $a<1,\;b>\frac{1}{2}\,(1+a)$}

Now $A>0$ and $B<A$.  Hence the self-gravity term gets dominant for large $k$
and makes $\omega^{2}$ negative.  For small $k$, $\omega^{2}$ is positive
since it is dominated by the pressure term ($B<0$) and/or the rotation term
($B\geq0$).  As the zero(s) of $\omega^{2}(k)$ can only be determined
numerically, case by case, we do not give a stability criterion but note that
the disc is \emph{unstable at small scales}, like a cold non-turbulent disc.

\subsection{Case $a>1,\;b<\frac{1}{2}\,(1+a)$}

In contrast to the previous case, $A<0$ and $B>A$.  So $\omega^{2}$ is
dominated by the self-gravity term and is negative for small $k$, while it is
dominated by the pressure/rotation term and is positive for large $k$.  Thus
the disc is \emph{unstable at large scales}, like a non-rotating
non-turbulent sheet.

\subsection{Case $a<1,\;b<\frac{1}{2}\,(1+a)$}

When $0<A<B$, the response of the disc is driven by pressure at small scales
and by rotation at large scales, while self-gravity acts more strongly at
intermediate scales.  Therefore this is a Toomre-like case: $\omega^{2}(k)$
has a minimum, which determines whether the disc is stable or not.  More
precisely, the minimum of $\omega^{2}(k)$ provides three useful pieces of
information: the stability threshold, the most unstable scale and its growth
rate.  Such quantities are introduced below, while illustrative cases are
discussed in Sect.\ 4.

   \emph{The stability threshold} $\overline{Q}$ is the value of $Q$ above
which the disc is stable at all scales:
\begin{equation}
\mbox{STABLE\ }\forall k\Longleftrightarrow
Q\geq\overline{Q}\,.
\end{equation}
$\overline{Q}$ can be determined by imposing that the minimum of
$\omega^{2}(k)$ vanishes.  Even though the calculations are very lengthy, the
formula is simple, especially if expressed in terms of the `right'
parameters:
\begin{equation}
\overline{Q}=2\left[A^{A}B^{-B}\left(B-A\right)^{B-A}\left(\frac{\ell_{0}}{
\ell_{\mathrm{T}}}\right)^{2A-B}\right]^{1/(2A)}\,,
\end{equation}
where $\ell_{\mathrm{T}}=1/k_{\mathrm{T}}$ is the Toomre scale.  Eq.\ (16),
with $\overline{Q}$ specified by Eq.\ (17), is our stability criterion.  It
reduces to Toomre's criterion $Q\geq1$ in the limiting case of a
non-turbulent disc: $A=1,\;B=2$ ($a=b=0$).

   \emph{The most unstable scale}, $\ell_{\mathrm{min}}=1/k_{\mathrm{min}}$,
is given by the formula:
\begin{equation}
\frac{\ell_{\mathrm{min}}}{\ell_{\mathrm{T}}}=\overline{\mathcal{L}}\left(Q/
\,\overline{Q}\,\right)^{2/(B-A)}\,,
\end{equation}
where $Q/\,\overline{Q}$ measures the stability level of the disc (like $Q$
in Toomre's case), and $\overline{\mathcal{L}}$ is the value of
$\ell_{\mathrm{min}}/\ell_{\mathrm{T}}$ at the stability threshold
($Q/\,\overline{Q}=1$):
\begin{equation}
\overline{\mathcal{L}}=\left[\frac{B-A}{B}\left(\frac{\ell_{0}}{\ell_{\mathrm
{T}}}\right)^{A-1}\right]^{1/A}\,.
\end{equation}
Eq.\ (18) reduces to
$\ell_{\mathrm{min}}/\ell_{\mathrm{T}}=\frac{1}{2}\,Q^{2}$ in Toomre's case.

   \emph{The growth rate of the most unstable scale},
$\gamma_{\mathrm{min}}=(-\omega_{\mathrm{min}}^2)^{1/2}$, is given by the
formula:
\begin{equation}
\frac{\omega_{\mathrm{min}}^{2}}{\kappa^{2}}=1-\left(Q/\,\overline{Q}\,\right
)^{-2A/(B-A)}\,,
\end{equation}
which vanishes at the stability threshold.  Eq.\ (20) reduces to
$\omega_{\mathrm{min}}^{2}/\kappa^{2}=1-Q^{-2}$ in Toomre's case.

\subsection{Case $a>1,\;b>\frac{1}{2}\,(1+a)$}

Even $B<A<0$ is a Toomre-like case.  In fact, although the scales at which
pressure and rotation dominate are reversed, self-gravity still controls
intermediate scales and $\omega^{2}(k)$ has a minimum, which determines
whether the disc is stable or not.  This means that, even now, the stability
criterion is:
\begin{equation}
\mbox{STABLE\ }\forall k\Longleftrightarrow
Q\geq\overline{Q}\,.
\end{equation}
The stability threshold $\overline{Q}$, the most unstable scale and its
growth rate, $\ell_{\mathrm{min}}$ and $\gamma_{\mathrm{min}}$, are given by
the same formulae as in Eqs (17)--(20).

\begin{figure*}
\includegraphics[angle=-90.,scale=.99]{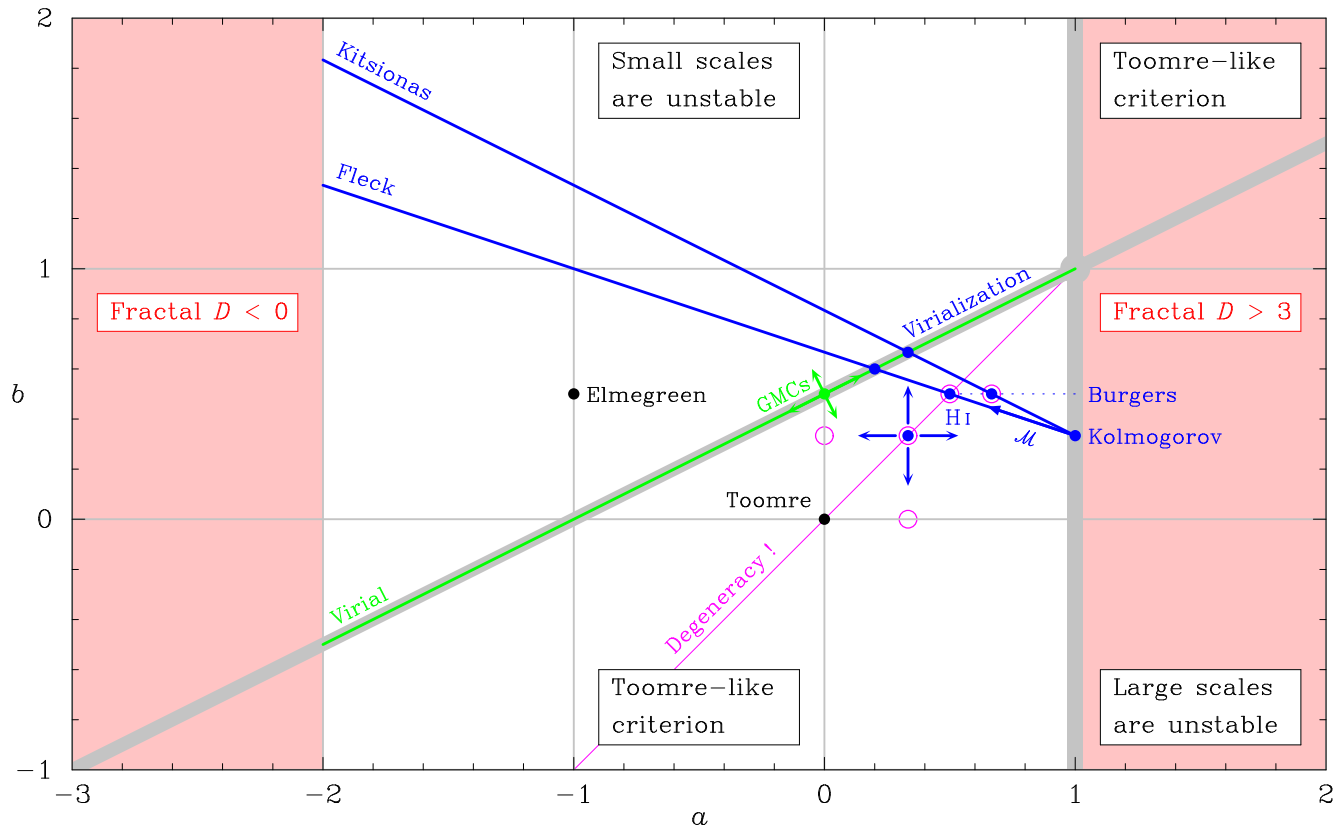}
\caption{The stability map of turbulence.  The coordinates $a$ and $b$ are
         the exponents of the density- and velocity-size scaling relations
         [see Eqs (3) and (4)].  The shaded lines, their intersection point
         and the regions between them represent the variety of stability
         regimes possessed by clumpy/turbulent discs.  The points
         $(a,b)=(0,0)$ and $(a,b)=(-1,\frac{1}{2})$ correspond to the
         limiting case of a non-turbulent disc and to the case investigated
         by Elmegreen (1996).  The non-shaded part of the plane shows the
         natural range of $a$.  The thick lines are phenomenological models
         of GMC and H\,\textsc{i} turbulence, the semiclosed-head arrows
         represent observations, while the closed-head arrow represents
         simulations with increasing $\mathcal{M}$ach number.  The solid
         circles correspond to points of special interest in astrophysical
         turbulence.  Also shown are the cases illustrated in Figs 2 and 3
         (hollow circles), and the degeneracy condition discussed in Sect.\ 4
         (thin line).  Note that the stability criterion is $Q\geq1$ along
         such a line.}
\end{figure*}

\section{THE STABILITY MAP OF TURBULENCE}

The results of Sect.\ 2 are summarized in Fig.\ 1.  The $(a,b)$ plane is
divided into four regions, where stability \`{a} la Toomre alternates with
instability at small or large scales.  The two shaded lines that separate
those regions, and the point at which those lines intersect, represent
transitions between different stability phases.  Thus the corresponding
stability criteria are hybrid (see Sects 2.2--2.4).  The points $(a,b)=(0,0)$
and $(a,b)=(-1,\frac{1}{2})$ represent the limiting case of a non-turbulent
disc and the case investigated by Elmegreen (1996).  To the best of our
knowledge, this was the only theoretical work devoted to the gravitational
instability of clumpy/turbulent discs before ours.  Fig.\ 1 also illustrates
the most interesting stability regimes populated by models, simulations and
observations of astrophysical turbulence.  Such points are discussed
throughout the rest of this section.  Specific predictions about strongly and
weakly supersonic turbulence are then made in Sect.\ 3.1.

   The mass-size scaling relation, $M\propto\Sigma\ell^{2}\propto\ell^{a+2}$,
tells us the natural bounds of $a$.  In fact, $D=a+2$ is the fractal
dimension of the mass distribution, which ranges from 0 to 3, so we have
$-2\leq a\leq1$.  Note that the upper bound corresponds to the case in which
the destabilizing effect of self-gravity is scale-independent, i.e.\ to the
vertical shaded line introduced above.

   Does even the other shaded line have a twofold meaning?  Yes, and an
important one!  If the stabilizing effect of pressure has the same
scale-dependence as the effect of self-gravity, $b=\frac{1}{2}\,(1+a)$, then
$\sigma^{2}\propto\ell\Sigma\propto M/\ell$, which is the virial scaling
relation.  GMCs are then expected to clump along that line, i.e.\ to populate
the transition regime between stability \`{a} la Toomre and instability at
small scales.  For example, the well-known scaling laws $\Sigma=constant$ and
$\sigma\propto\ell^{1/2}$ (Larson 1981; Solomon et al.\ 1987) correspond to
the point $(a,b)=(0,\frac{1}{2})$.  Both Galactic and extragalactic GMCs show
non-negligible dispersion around that point, especially along the virial
line, as can be inferred from Bolatto et al.\ (2008) and Heyer et al.\
(2009).  The case of perturbed galactic environments seems different.
Rosolowsky \& Blitz (2005) investigated the physical properties of GMCs in
M64 (NGC 4826), an interacting molecule-rich galaxy, and found $\Sigma\propto
M^{0.7\pm0.2}$ and $\sigma\propto\ell^{1.0\pm0.3}$, which means
$(a,b)=(5^{+13}_{-3},1.0\pm0.3)$.  If such scaling relations apply to
individual GMCs, as was originally suggested, then each cloud is far from
being in simple virial equilibrium.  Besides, since $(a,b)$ is below the
virial line and on the right of the $a=1$ line, the $\mathrm{H}_{2}$ disc is
unstable at large scales (in the sense specified in Sect.\ 2.6) and the
fractal dimension is formally higher than three.  Alternatively, one may
argue that those scaling relations arise from the superposition of more GMCs,
each one characterized by the standard scaling laws, but with proportionality
factors varying significantly over the disc (Rosolowsky, private
communication).

   Now what about H\,\textsc{i}\,?  A turbulence model that is becoming more
and more popular is the one introduced by Fleck (1996), which predicts
$\rho^{1/3}\sigma\propto\ell^{1/3}$.  To understand the meaning of this
scaling relation, compare it with Kolmogorov's law $\sigma\propto\ell^{1/3}$.
Fleck's relation tells us that in a turbulent medium with both velocity and
density fluctuations the quantity $\rho^{1/3}\sigma$ plays a role similar to
$\sigma$ in incompressible turbulence.  Fleck (1996) assumed that
$\Sigma\sim\rho\ell$, which means $\ell\la h$ (see Sect.\ 2.1).  So his
prediction corresponds to the line $b=\frac{1}{3}\,(2-a)$, which crosses
several stability regimes.  The limiting case of Kolmogorov turbulence,
$(a,b)=(1,\frac{1}{3})$, lies in the transition regime between stability
\`{a} la Toomre and instability at large scales.  Both high-resolution
simulations of supersonic turbulence and H\,\textsc{i} observations populate
the Toomre-like domain.  Such simulations cluster along the Fleck line, near
$(a,b)=(\frac{1}{2},\frac{1}{2})$, the case of Burgers turbulence (Kowal \&
Lazarian 2007; Kritsuk et al.\ 2007; Schmidt et al.\ 2008; Price \& Federrath
2010; see also Fleck 1996 and references therein).  In weakly supersonic
regimes, simulations cluster closer to the Kolmogorov limit
$(a,b)=(1,\frac{1}{3})$, as we will show in Sect.\ 3.1.  Observed
H\,\textsc{i} intensity fluctuations, which are primarily due to \emph{cold}%
\footnote{Hereafter we will omit `cold' when referring to H\,\textsc{i},
          since our analysis also focuses on the cold ISM (see Sect.\ 2.1).}
H\,\textsc{i} (Lazarian \& Pogosyan 2000), show large scatter around
$(a,b)=(\frac{1}{3},\frac{1}{3})$, i.e.\ suggest a Kolmogorov scaling for
both velocity and density fluctuations (e.g., Lazarian \& Pogosyan 2000;
Elmegreen et al.\ 2001; Begum et al.\ 2006; Dutta et al.\ 2008, 2009a,\,b).
Such a scaling is also consistent with other H\,\textsc{i} observations
(e.g., Kim et al.\ 2007; Roy et al.\ 2008).  The simulations by Agertz et
al.\ (2009a), designed to explore the development of H\,\textsc{i} turbulence
in disc galaxies, suggest power-law indices consistent with the observed
ones, except before the fragmentation of the disc (Agertz, Romeo et al., in
preparation).

   \emph{Note} that there is a very interesting and unexpected case where
Fleck's model fits the observations well: the transition from H\,\textsc{i}
to GMC turbulence.  Clumps that climb up the Fleck line become progressively
more self-gravitating and will virialize at
$(a,b)=(\frac{1}{5},\frac{3}{5})$.  This is close to $(a,b)=(0,\frac{1}{2})$,
the reference point for GMC observations.  The predicted energy spectra are
$E_{\Sigma}(k)\propto\mathrm{d}\Sigma^{2}/\mathrm{d}k\propto k^{-7/5}$ and
$E_{\sigma}(k)\propto\mathrm{d}\sigma^{2}/\mathrm{d}k\propto k^{-11/5}$ (the
Kolmogorov spectrum scales as $k^{-5/3}$).

   Low-resolution simulations of supersonic turbulence suggest an alternative
scaling relation: $\rho^{1/2}\sigma\propto\ell^{1/3}$ (Kitsionas et al.\
2009).  In comparison with the Fleck case, this scaling gives more weight to
density fluctuations and translates into a steeper line:
$b=\frac{1}{2}\,(\frac{5}{3}-a)$, where it is again assumed that
$\Sigma\sim\rho\ell$.  The cases of Kolmogorov and Burgers turbulence
correspond to $(a,b)=(1,\frac{1}{3})$ and $(a,b)=(\frac{2}{3},\frac{1}{2})$,
while the virialization point is $(a,b)=(\frac{1}{3},\frac{2}{3})$.  The
Kitsionas line crosses the same stability regimes as the Fleck line, but lies
farther away from H\,\textsc{i} and GMC observations.

\subsection{Strongly vs.\ weakly supersonic turbulence}

Although our analysis focuses on strongly supersonic turbulence (see Sect.\
2.1), here we extend it to weakly supersonic regimes (the case of a transonic
or subsonic medium was considered in Sect.\ 2.1).

   How does the Mach number affect the density- and velocity-size scaling
relations?  And how does it affect the stability of the disc?  To answer
these questions, one should not compute $a$ and $b$ directly from the density
and velocity spectral indices.  One should first evaluate the typical density
contrast of the medium (see Sect.\ 2.1).  The density probability
distribution is approximately lognormal, and its mean $\mu$ and standard
deviation $\mathrm{SD}$ depend on the rms Mach number $\mathcal{M}$ (Padoan
et al.\ 1997): $\mu=\frac{1}{2}\mathrm{SD}^{2}
\approx\frac{1}{2}\ln(1+\frac{1}{4}\mathcal{M}^{2})$.  For such a
distribution, the mass-weighted median density is
$\rho_{\mathrm{med}}=\bar{\rho}\,\mathrm{e}^{\mu}
\approx\bar{\rho}\,(1+\frac{1}{4}\mathcal{M}^{2})^{1/2}$, where $\bar{\rho}$
is the mean density (see sect.\ 2.1.4 of McKee \& Ostriker 2007).  This
provides a robust estimate of the typical density (mean plus fluctuations) in
the medium.  The corresponding density contrast is
$\delta_{\mathrm{med}}=(\rho_{\mathrm{med}}-\bar{\rho})/\bar{\rho}$.

   In weakly supersonic turbulence ($\mathcal{M}\approx2$), we have
$\delta_{\mathrm{med}}\approx0.4$ so the mass column density at scale $\ell$
is dominated by the mean density:
$\Sigma_{\mathrm{eff}}\approx\bar{\rho}\ell$ (hence $a\approx1$) for $\ell\la
h$, while $\Sigma_{\mathrm{eff}}\approx\bar{\rho}h$ (hence $a\approx0$) for
$\ell\ga h$.  In contrast, the 1D velocity dispersion at scale $\ell$ is
dominated by the turbulent term:
$\sigma\approx\sigma_{\mathrm{tur}}(\ell)\propto\ell^{b}$.  For $\ell\la h$,
we can relate $b$ to $a$ using Fleck's model, $b=\frac{1}{3}\,(2-a)$, and get
$(a,b)\approx(1,\frac{1}{3})$.  \emph{Thus} weakly supersonic 3D turbulence
drives the disc to the boundary of the Toomre-like domain, near the
Kolmogorov point.

   In strongly supersonic turbulence ($\mathcal{M}\ga5$), we have
$\delta_{\mathrm{med}}\ga2$ so both the mass column density and the 1D
velocity dispersion are dominated by turbulent fluctuations.  We can then
compute $a$ and $b$ from the density and velocity spectral indices (see
Sect.\ 2.1).  Simulations show that in an isothermal medium with negligible
self-gravity, with or without magnetic fields, the density spectrum flattens
(Beresnyak et al.\ 2005; Kim \& Ryu 2005; Kowal et al.\ 2007) and the
velocity spectrum steepens (e.g., Kowal \& Lazarian 2007; Kritsuk et al.\
2007; Price \& Federrath 2010) as the Mach number increases.  At Mach 7, we
find $(a,b)\sim(0.8,0.4)$ from Kowal et al.\ (2007) and $(a,b)\sim(0.3,0.6)$
from Kowal \& Lazarian (2007), using Fleck's model in both cases.%
\footnote{In the first case, $\mathrm{d}\ln E_{\rho}/\mathrm{d}\ln k$ (i.e.\
          $-r$) is given in table 2 of Kowal et al.\ (2007); $a$ is computed
          from $r$ as in Sect.\ 2.1 ($\ell\la h$): $a=\frac{1}{2}\,(r+1)$;
          and $b$ is computed from $a$ using Fleck's model:
          $b=\frac{1}{3}\,(2-a)$.  This yields
          $(a,b)=(0.75\pm0.05,0.42\pm0.02)$ for the sub-Alfv\'{e}nic
          simulation and $(a,b)=(0.80\pm0.10,0.40\pm0.03)$ for the
          super-Alfv\'{e}nic simulation.  In the second case (of the main
          text), $\mathrm{d}\ln E_{v}/\mathrm{d}\ln k$ (i.e.\ $-s$) is given
          in fig.\ 1 of Kowal \& Lazarian (2007); $b$ is computed from $s$ as
          in Sect.\ 2.1: $b=\frac{1}{2}\,(s-1)$; and $a$ is computed from $b$
          using Fleck's model: $a=2-3b$.  This yields
          $(a,b)=(0.29\pm0.08,0.57\pm0.03)$.  This simulation is
          sub-Alfv\'enic, like the first one above.}
In spite of the large uncertainties, it is clear that these points are on the
left of $(1,\frac{1}{3})$ and, on average, closer to
$(\frac{1}{2},\frac{1}{2})$.  \emph{Thus} strongly supersonic 3D turbulence
drives the disc well inside the Toomre-like domain, near the Burgers point.

\begin{figure}
\includegraphics[angle=-90.,scale=1.01]{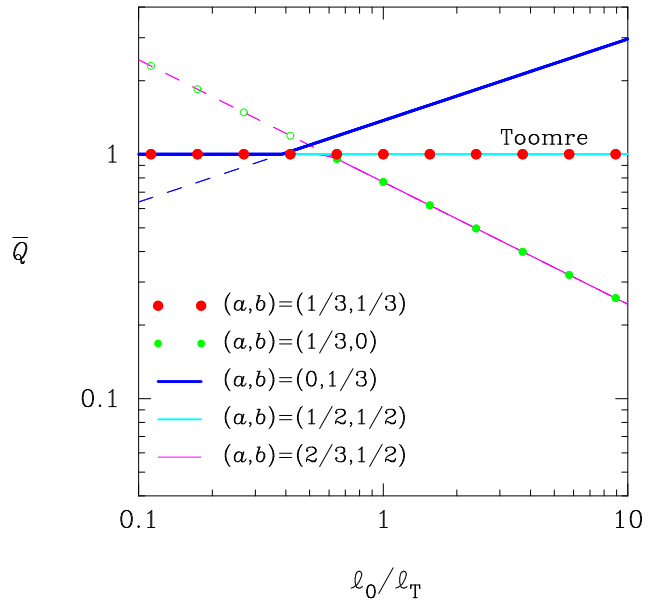}
\caption{The stability threshold of clumpy/turbulent discs, $\overline{Q} =
         \overline{Q}(a,b,\ell_{0})$, where $a$ and $b$ are the exponents of
         the density- and velocity-size scaling relations, and $\ell_{0}$ is
         the typical scale at which density and velocity saturate.  In
         addition, $\ell_{\mathrm{T}}$ is the Toomre scale.  The dashed lines
         and the hollow circles show the power-law behaviour predicted by
         Eq.\ (17) when density and velocity do not saturate; $\ell_{0}$ is
         then the fiducial scale at which those quantities are observed.  The
         limiting case of a non-turbulent disc is $\overline{Q}=1$.}
\end{figure}

\begin{figure}
\includegraphics[angle=-90.,scale=.99]{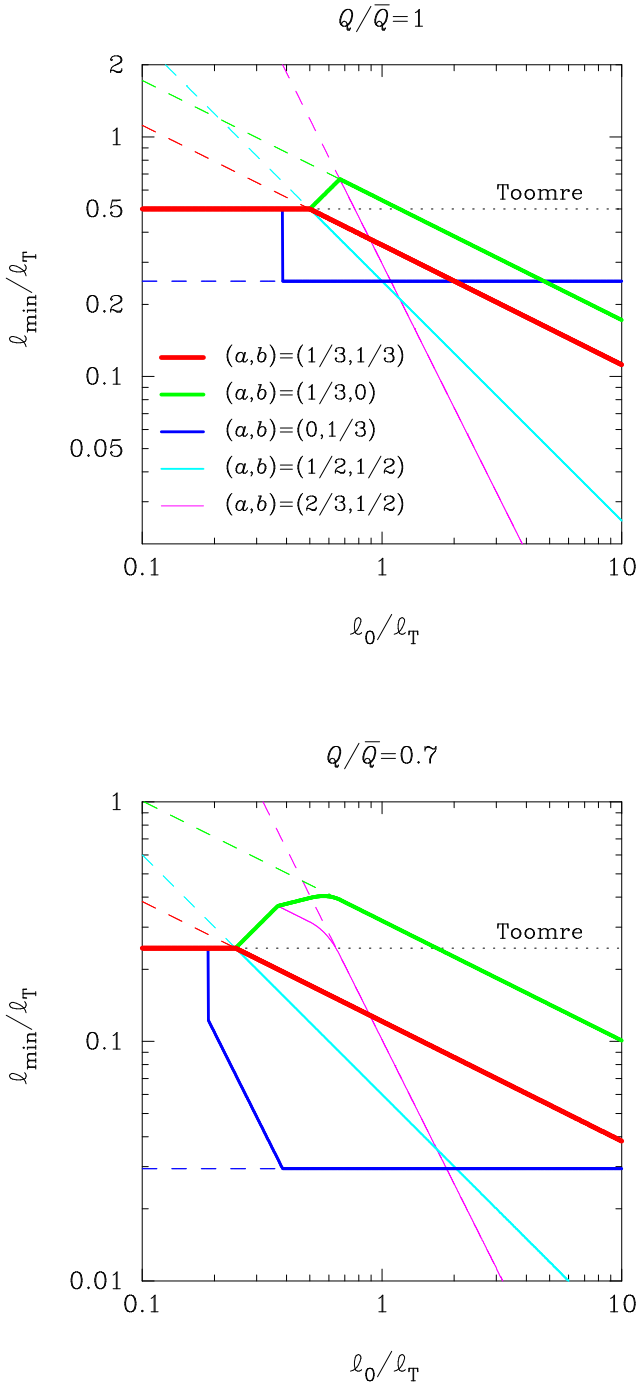}
\caption{The most unstable scale of clumpy/turbulent discs,
         $\ell_{\mathrm{min}} =
         \ell_{\mathrm{min}}(a,b,\ell_{0},Q/\,\overline{Q})$, where $a$ and
         $b$ are the exponents of the density- and velocity-size scaling
         relations, $\ell_{0}$ is the typical scale at which density and
         velocity saturate, and $Q/\,\overline{Q}$ is the stability level.
         In addition, $\ell_{\mathrm{T}}$ is the Toomre scale.  The dashed
         lines show the power-law behaviour predicted by Eqs (18) and (19)
         when density and velocity do not saturate; $\ell_{0}$ is then the
         fiducial scale at which those quantities are observed.  Also shown
         is the limiting case of a non-turbulent disc (dotted line).}
\end{figure}

\section{ILLUSTRATIVE CASES}

Let us now illustrate the stability characteristics of clumpy/turbulent discs
in a number of cases, those marked with hollow circles in the Toomre-like
domain of our map (see Fig.\ 1).  The points $(\frac{1}{3},\frac{1}{3})$,
$(\frac{1}{2},\frac{1}{2})$ and $(\frac{2}{3},\frac{1}{2})$ are typical
values of $(a,b)$ inferred from H\,\textsc{i} observations, high- and
low-resolution simulations of supersonic turbulence.  The points
$(\frac{1}{3},0)$ and $(0,\frac{1}{3})$ are the contributions of density and
velocity fluctuations to the observed $(a,b)=(\frac{1}{3},\frac{1}{3})$.  For
each case, we proceed in two ways.  First, we assume that the density- and
velocity-size scaling relations are perfect power laws, as given by Eqs (3)
and (4), so that $\ell_{0}=1/k_{0}$ is the fiducial scale at which density
and velocity are observed.  We then evaluate the stability characteristics
analytically using Eqs (17)--(20).  Second, we consider more realistic
scaling relations, which take into account the saturation of density and
velocity at large scales:
\begin{equation}
\Sigma_{\mathrm{eff}}=\Sigma_{0}\,\mathcal{D}_{\ell}\,,\;\;\;\;\;
\mathcal{D}_{\ell}=\left\{\begin{array}{ll}
                          (\ell/\ell_{0})^{a}
                            & \mbox{if\ }\ell\leq\ell_{0}\,, \\
                          1
                            & \mbox{else}\,;
                          \end{array}
                   \right.
\end{equation}
\begin{equation}
\sigma=\sigma_{0}\,\mathcal{V}_{\ell}\,,\;\;\;\;\;
\mathcal{V}_{\ell}=\left\{\begin{array}{ll}
                          (\ell/\ell_{0})^{b}
                            & \mbox{if\ }\ell\leq\ell_{0}\,, \\
                          1
                            & \mbox{else}\,;
                          \end{array}
                   \right.
\end{equation}
where $\ell_{0}$ is now the typical saturation scale.  We then evaluate the
stability characteristics numerically using the dispersion relation, Eq.\
(2), which we rewrite as:
\begin{equation}
\frac{\omega^{2}}{\kappa^{2}}=1-\frac{\mathcal{D}_{\ell}}{(\ell/\ell_{\mathrm
{T}})}+\frac{Q^{2}}{4}\,\frac{\mathcal{V}_{\ell}^{2}}{(\ell/\ell_{\mathrm{T}}
)^{2}}\,,
\end{equation}
where $Q=\kappa\sigma_{0}/\pi G\Sigma_{0}$ is the Toomre parameter and
$\ell_{\mathrm{T}}=2\pi G\Sigma_{0}/\kappa^{2}$ is the Toomre scale.

   Fig.\ 2 shows the stability threshold $\overline{Q}$, i.e.\ the value of
$Q$ above which the disc is stable at all scales.  The first curious result
is that such a diagnostic is highly degenerate.  For example, look at the
cases $(a,b)=(\frac{1}{3},\frac{1}{3})$ and
$(a,b)=(\frac{1}{2},\frac{1}{2})$, which represent H\,\textsc{i} observations
and high-resolution simulations of supersonic turbulence.  They have
$\overline{Q}\equiv1$!  Why do such cases degenerate into Toomre's case?  Why
doesn't turbulence show up?  Eq.\ (17) gives us the answer: because the
effects of density and velocity fluctuations cancel out when $2A-B=0$, i.e.\
along the line $b=a$ (see the map).  The cases $(a,b)=(\frac{1}{3},0)$ and
$(a,b)=(0,\frac{1}{3})$ allow us to disentangle such effects.  Density
fluctuations that saturate at a typical scale $\ell_{0}$ tend to stabilize
the disc by decreasing the stability threshold: $\overline{Q}<1$ if
$\ell_{0}\ga\frac{1}{2}\ell_{\mathrm{T}}$, and $\overline{Q}=1$ otherwise.
To understand this result, remember that such fluctuations reduce the
self-gravity term in the dispersion relation by a factor $\mathcal{D}_{\ell}$
[see Eqs (22) and (24)], and that self-gravity is destabilizing.  Velocity
fluctuations have an antagonistic effect.  They reduce pressure by a factor
$\mathcal{V}_{\ell}^{2}$ [see Eqs (23) and (24)], and tend to destabilize the
disc by increasing the stability threshold: $\overline{Q}>1$ if
$\ell_{0}\ga\frac{1}{2}\ell_{\mathrm{T}}$, and $\overline{Q}=1$ otherwise.
When density/velocity fluctuations do not saturate, their effect becomes
destabilizing/stabilizing if the fiducial scale is small.

   Fig.\ 3 shows the most unstable scale, $\ell_{\mathrm{min}}$, for two
values of $Q/\,\overline{Q}$.  Remember that this quantity measures the
stability level of the disc, like $Q$ in Toomre's case.  So
$Q/\,\overline{Q}=1$ (top panel) means that the disc is marginally unstable,
while $Q/\,\overline{Q}=0.7$ (bottom panel) means that the disc is moderately
unstable.  In contrast to $\overline{Q}$, $\ell_{\mathrm{min}}$ is not
degenerate.  Turbulence now has a significant effect in the case of
H\,\textsc{i} observations, since the contributions of density and velocity
fluctuations are no longer antagonistic.  For
$\ell_{0}\sim\ell_{\mathrm{T}}$, $\ell_{\mathrm{min}}$ is about 30 to 50 per
cent smaller than in Toomre's case, depending on the value of
$Q/\,\overline{Q}$.  The impact of turbulence is stronger in the case of
high-resolution simulations: the most unstable scale is typically half an
order of magnitude below the expected value.  Turbulent effects become less
important at low resolution.

   The growth rate of the most unstable scale, $\gamma_{\mathrm{min}}$, is
independent of $\ell_{0}$ and vanishes for $Q/\,\overline{Q}=1$ [see Eq.\
(20)].  For $Q/\,\overline{Q}<1$, the effects of density and velocity
fluctuations cancel out when $A/(B-A)=1$, i.e.\ $b=a$.  This degeneracy
condition is the same as that found for $\overline{Q}$, and has the same
consequences.

\section{CONCLUSIONS}

\begin{itemize}
\item Observations and simulations of the interstellar medium are revealing
      its turbulent nature with higher and higher fidelity.  Such information
      must then be taken into account when analysing the stability of
      galactic discs.  Our contribution is a natural extension to Toomre's
      work, which will prove useful for both low- and high-redshift analyses.
\item Turbulence has a deep impact on the gravitational instability of the
      disc.  It excites a rich variety of stability regimes, several of which
      have no classical counterpart.  We illustrate this result in the form
      of a map, which relates our stability scenario to the phenomenology of
      interstellar turbulence: GMC/H\,\textsc{i} observations, simulations
      and models.
\item GMC turbulence drives the disc to a regime of transition between
      instability at small scales and stability \`{a} la Toomre.  Toomre's
      criterion works instead typically well when applied to discs of cold
      H\,\textsc{i}, since the effects of density and velocity fluctuations
      tend to cancel out.  Even so, H\,\textsc{i} turbulence produces a clear
      signature in disc morphology.  It reduces the characteristic scale of
      instability by 30--50 per cent or more, depending on the value of $Q$
      and on the shape of the energy spectrum.  The transition from
      H\,\textsc{i} to GMC turbulence occurs when $\Sigma\sim\ell^{1/5}$ and
      $\sigma\sim\ell^{3/5}$ (for more information see Sect.\ 3).
\item Coming astronomical facilities such as ALMA%
\footnote{http://www.eso.org/sci/facilities/alma/}
      will be able to resolve the scaling properties of galactic turbulence
      up to very high redshifts.  Using our map, such data will show up as
      \emph{evolutionary tracks}, which will reveal the interplay between
      gravitational instability and turbulence during the galaxy life.  In
      turn, this will be useful for constraining the sources of galactic
      turbulence and for understanding the evolution of cosmic star
      formation.
\end{itemize}

\section*{ACKNOWLEDGMENTS}

We are very grateful to John Black, Leo Blitz, Rick Hessman, Ben Moore and
Erik Rosolowsky for useful discussions.  We are also grateful to an anonymous
referee for constructive comments and suggestions.  The first author thanks
Bruce Elmegreen for strong encouragement and valuable feedback on the
occasion of the IAU Symposium 254 `The Galaxy Disk in Cosmological Context'.
He also thanks the warm hospitality of both the Department of Physics at the
University of Gothenburg and the Department of Fundamental Physics at
Chalmers.

\bsp

\label{lastpage}

\end{document}